\documentclass[12pt,preprint]{aastex63}
\DeclareUnicodeCharacter{2212}{\ensuremath{-}}

\shorttitle{Radio Bursts In Eruptive \& Confined Flares}
\shortauthors{White et al.}

\begin{document}

\title{The Association of Solar Radio Bursts with Eruptive and Confined Flares}

\author{Stephen M. White}
\email{stephen.white.24@us.af.mil}
\affiliation{Space Warfare Directorate, Air Force Research Laboratory, Kirtland AFB, NM, USA}
\affiliation{Department of Physics and Astronomy, University of New Mexico, Albuquerque, NM 87106, USA}

\author{Maria D. Kazachenko}
%\email{}
\affiliation{Dept. of Astrophysical and Planetary Sciences, University of Colorado,  Boulder, CO 80305, USA}
\affiliation{Laboratory for Atmospheric and Space Physics, University of Colorado, Boulder, CO 80303, USA}
\affiliation{National Solar Observatory, 3665 Discovery Drive, Boulder, CO 80303, USA} 

\author{Edward W. Cliver}
%\email{}
\affiliation{National Solar Observatory, 3665 Discovery Drive, Boulder, CO 80303, USA}

\email{Draft: \today}

\begin{abstract}
We classify the metric radio emission associated with eruptive (CME-associated) and confined flares using the sample of events
between 2010 and 2016 identified by Kazachenko (2023). We find striking differences in the occurrence of radio bursts between the two classes of flare: for soft X-ray flare sizes above M1.4, confined flares largely lack Type II (1\% association rate) and Type IV (6\%) emission. Approximately 15\% of the sample of $\geq$M1.4 confined flares are associated with impulsive-phase Type III bursts. On the other hand, eruptive
flares are associated with Type II, Type III (both impulsive and late phase), and Type IV
bursts 45-60\% of the time. The different types of radio burst are associated with different drivers (IIIs with electron beams, IIs with shocks, IVs with post-flare loops), so it is striking that the connection of all three burst types to eruptive flares is so pronounced. These results can be interpreted in terms of the reconnection
topology of the principle candidate flare types, viz., reconnection between closed field lines for
confined flares and X-point reconnection in a CSHKP model for eruptive flares, with
interchange reconnection for jet-type flares and Type III bursts. 
\end{abstract}


\section{Introduction}

The classical metric solar radio burst types identified early in the history of radio astronomy \citep[e.g.][]{WSW63} still remain somewhat of a mystery. While we know the phenomena that produce them, at least in some cases, and the basics of the (coherent) emission mechanism by which they radiate, we still don't have a clear picture of all the details involved. Thus, we know that Type III bursts are produced by beams of electrons propagating out of the corona on open field lines, but the mechanism generating the electron beams and their role in solar flares remains a puzzle. We know that Type II bursts must be associated with shocks, but the nature of the shock source is not clear: coronal mass ejections (CMEs) are the obvious candidate driver for shocks, but there are many Type II bursts that occur with no detectable CME present. Type IV bursts are broadband features that can last for hours in the post-flare period, and so might be associated with post-flare loops, but they are rare enough that some additional specific condition must be present in the loops to produce the observed radio emission.

A prominent reason for the lack of detailed understanding is the fact that we do not detect the actual drivers of the radio bursts at any other wavelength. These bursts are dominant in the frequency range below 200 MHz, in regions of the corona where electron densities $n_e$ are below 10$^9$ cm$^{-3}$: we know this because they emit via the coherent plasma emission mechanism, at the ambient plasma frequency $f_p\,=\,9000\sqrt{n_e}$ Hz and its harmonic. However, the densities of the nonthermal, typically keV-energy particles that actually produce the radio emission are much lower than this, and while, e.g. Type III electron beams are readily detected in situ in the solar wind by satellites, they do not produce emission at other wavelengths strong enough to be detected remotely. An additional complication for identification of matching features is the fact that the spatial resolution achievable in practice below 200 MHz is limited to tens of arcseconds, due both to instrumental limitations and to the presence of scattering in the solar atmosphere. Such spatial resolution is too poor for comparison with the arcsecond resolution available to identify coronal features at UV and EUV wavelengths. 

Despite this lack of detailed understanding, the presence or absence of such bursts in solar flares is still telling. In the case of Type III bursts, they typically occur in the onset of the impulsive phase, indicating that open field lines play a role in the initial energy release in the associated flare. Not all flares exhibit Type II bursts, and they are readily detected from smaller B- and C-class X-ray flares, indicating that flare size is not a critical factor in the formation of the associated shocks. Type IV bursts have been linked in the past with solar energetic particle events. Thus, the occurrence of radio bursts in conjunction with different types of solar flare can reveal features that play a role in flare occurrence and development.

One of the primary distinctions in flare types is between eruptive and confined flares. Eruptive flares, which exhibit CMEs, filament eruptions and other indicators of large-scale disruption of the corona, are visually spectacular, and flares that have significant space weather impacts are generally eruptive. The ``standard'' flare model \citep[``CSHKP'':][]{Car64,Stu66,Hir74,KoP76} of reconnection at a current sheet necessarily involves an upwards outflow due to the reconnection exhaust, and is most commonly presented in conjunction with an eruption. Confined flares show no evidence for an eruption, but otherwise they are just as capable as eruptive flares of producing strong hard X-ray, soft X-ray, H$\alpha$ and microwave emission. In this paper we investigate the properties of the classical solar radio burst types associated with the classes of eruptive and confined flares using the catalog assembled by \citet{Kaz23}.

\section{Eruptive and confined solar flares}

\citet{Kaz23} presented a catalog of the features of eruptive and confined events that
is a valuable curated resource for studies of the comparative properties of the
two classes of flare \citep[e.g.][]{LRW24,CKH25}. The focus of that work was on comparing
thermodynamic and magnetic properties, such as thermal energy, active
region magnetic content and magnetic reconnection rates. The full catalog
consists of all flares of GOES class C5.0 and larger\footnote{The GOES soft X-ray (SXR) flare classification system is defined as follows: GOES classes A1-9
through X1-9 correspond to flare peak $1-8$ \AA\ fluxes of (1-9) $\times$ 
10$^n$ W m$^{-2}$ where $n$ = [−8, −7, −6, −5, −4] for classes A, B, C,
M, and X, respectively. The catalog identifies flare events with the
original reported flare classes using the legacy scaling of GOES XRS
soft X-ray emission. However, the catalog also includes peak GOES XRS
$1-8$ \AA\ fluxes and flare classes on the recently updated true physical scale, which gives values a factor $0.7^{-1}$ larger than the legacy scale. The figures in this paper report 
peak fluxes using the modern scale.} observed by the Solar Dynamics Observatory (SDO)
from 2010 to 2016 within 45\arcdeg\ of central meridian. Deriving magnetic parameters of the flaring active regions is an emphasis of the study, hence the avoidance of flares closer to the limb. Attribution of
the confined or eruptive quality of each flare was determined from the
database of \citet{LCH21}, which used a combination of CME detections in
the CDAW CME catalog \citep{YGM04} and EUV wave detections from the Atmospheric Imaging Assembly (AIA) on SDO. This results in an unbalanced sample with 152 eruptive and 328
confined flares. Smaller flares are significantly more likely to be
confined than eruptive, so a more balanced subsample was obtained by 
limiting the threshold to $>$M1.0 (on the legacy scale; M1.4 on the modern scale): this results in 103 eruptive and 101 confined flares. 

However, \citet{Kaz23} found that the main conclusions drawn from the study 
of the smaller balanced sample were mirrored by the larger catalog. Eruptive
flares have the same amount of reconnected flux as confined flares, but
they tend to occur in smaller active regions with weaker field strengths.
Confined flares have more compact flare ribbons and stronger field
strengths, reach higher temperatures and are more efficient in delivering 
flare energy in the form of accelerated particles. \citet{Kaz23}
attributes these results to reconnection in eruptive flares occurring
more slowly in larger current sheets higher in the corona where the 
magnetic field is weaker, while confined flares have reconnection in
more compact current sheets lower in the atmosphere.

\begin{figure}
\includegraphics[scale=0.62,keepaspectratio=true,angle=0,clip=false]{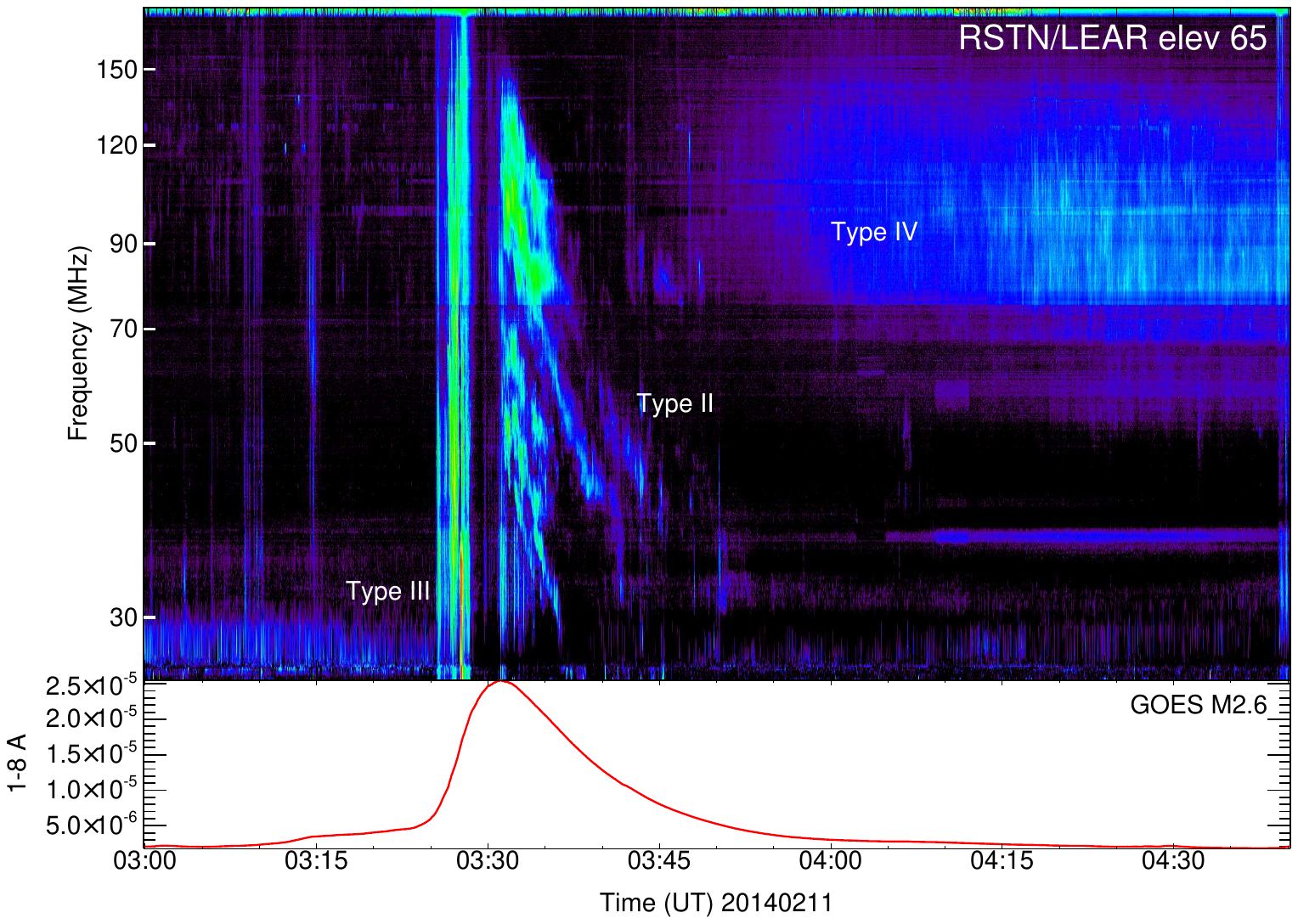}
\caption{A classic event showing Type II, III and IV bursts in the ``prototypical'' pattern: fast-drift Type III bursts starting early in the rise phase of the flare soft X-ray emission, a split-band Type II burst with both fundamental and harmonic traces drifting more slowly downwards in frequency from above 100 MHz to the bottom of the dynamic spectrum, and relatively broadband Type IV emission starting well after the Type II burst, lasting for an hour or more with very slow if any frequency drift. The data are from the Learmonth RSTN observatory on 2014 February 11. The lower panel shows the GOES XRS 1 - 8 \AA\ soft X-ray flux measurements on the same time axis as the radio data.\label{fig234}}
\end{figure}

This difference in properties can have other consequences. Thus, one
might expect that open field lines play a larger role in eruptive flares than
in confined flares, and \citet{DeB18} showed using global magnetic field
models (albeit for a highly unbalanced
sample) that confined events have less access to open field lines than
eruptive events. Type III solar radio bursts often occur early in the
impulsive phase of flares and are produced by electron beams
generated in the corona that propagate outwards on open field lines. If
eruptive events are initiated higher in the corona with weaker magnetic
field strengths, the Alfv{\'e}n speed is likely lower, resulting in quicker
formation of shocks \citep[e.g.][]{GLK01}. Such shocks can produce Type
II solar radio bursts, so it is possible that the difference between
eruptive and confined flares is reflected in the properties of associated 
Type II bursts (such as starting frequency).

\begin{figure}[t]
\centering
\includegraphics[scale=0.62,keepaspectratio=true,angle=0,clip=false]{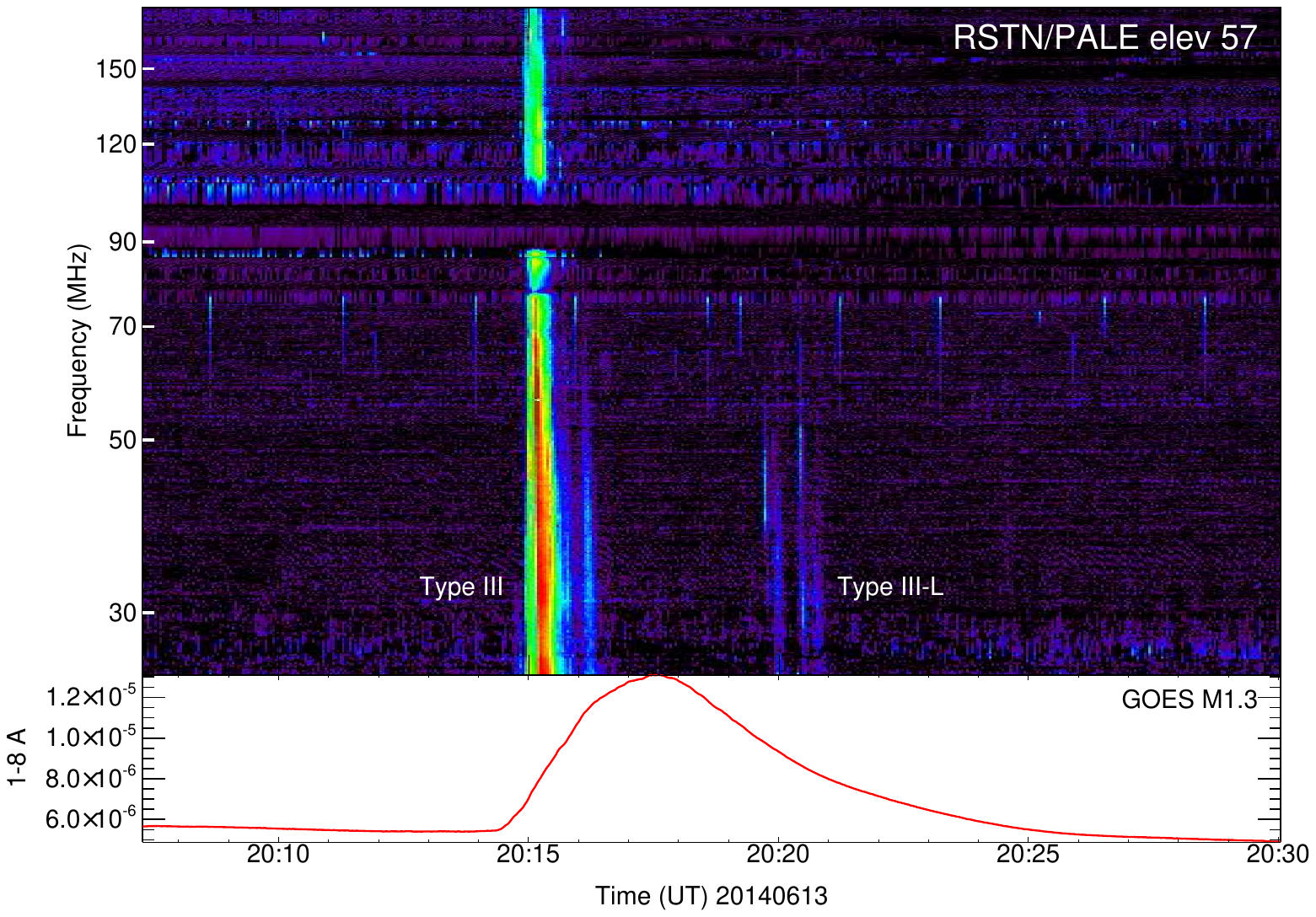}
\caption{The radio emission from the M1.3 flare on 2014 June 13. This flare exhibits the impulsive phase Type III burst during flare onset as well as a weak Type III during the decay phase, which we label as a Type III ``late'' (``T3L''). \label{fig33}}
\end{figure}

The purpose of this short paper is to expand the survey of properties associated
with eruptive and confined events to include solar radio bursts. We do
so by visual inspection of radio dynamic spectra for every event in the
\citet{Kaz23} catalog for which we can find data. While there have been many studies of radio bursts associated with CMEs, they have generally not directly compared radio bursts for confined and eruptive flares. \citet{Mit21} looked at X flares from Cycles 23 and 24, and found that X flares lacking CMEs also lacked interplanetary (IP) Type III bursts (as identified in spacecraft measurements of radio spectra below 14 MHz), while 90\% of X flares with CMEs were also accompanied by IP Type IIIs. \citet{MiP25} extended this study to M flares, and found that 32\% of confined (i.e., not accompanied by a detectable CME) flares exhibited IP Type III bursts, compared to 50\% of eruptive flares, while only 1\% of confined flares produced IP Type II bursts compared to 7\% of eruptive flares.
\citet{SMK05} found that 75\% of the metric Type II radio bursts in 2000 could be  associated with CMEs. 
From a sample of 290 CMEs associated with X-ray flares larger than C1 (on the legacy scale) in the years 1997 to 2000, \citet{SMV11} found that 21\% of the CME/flare pairs were associated with Type II events. 
Using the radio burst reports from RSTN observers provided by NOAA/SWPC in their event reports from 1996 to 2019, \citet{KMK23} found that 15\% of metric Type II bursts could not be associated with a CME launched within 2 hours of the radio burst onset. Type IIs without discernible CMEs tended to have shorter durations and slower frequency drift rates than those accompanied by CMEs. \citet{KMK21} carried out a similar analysis for Type IV bursts in solar cycle 24, and found that 81\% of Type IVs could be associated with CMEs. In contrast, only 2\% of detected CMEs could be associated with Type IV bursts.
\citet{LDC24} compiled a catalog of metric Type II bursts in solar cycle 24 by visual inspection of dynamic spectra, and found 429 events, of which 73\% could be associated with CMEs within 1 hour of the radio burst.

\section{Solar Radio Bursts}

Five low-frequency solar radio burst types are discussed here \citep[for more detailed reviews see, e.g.,][]{WSW63,Kun65,McL85,PiV08}:

\begin{figure}[t]
\centering
\includegraphics[scale=0.62,keepaspectratio=true,angle=0,clip=false]{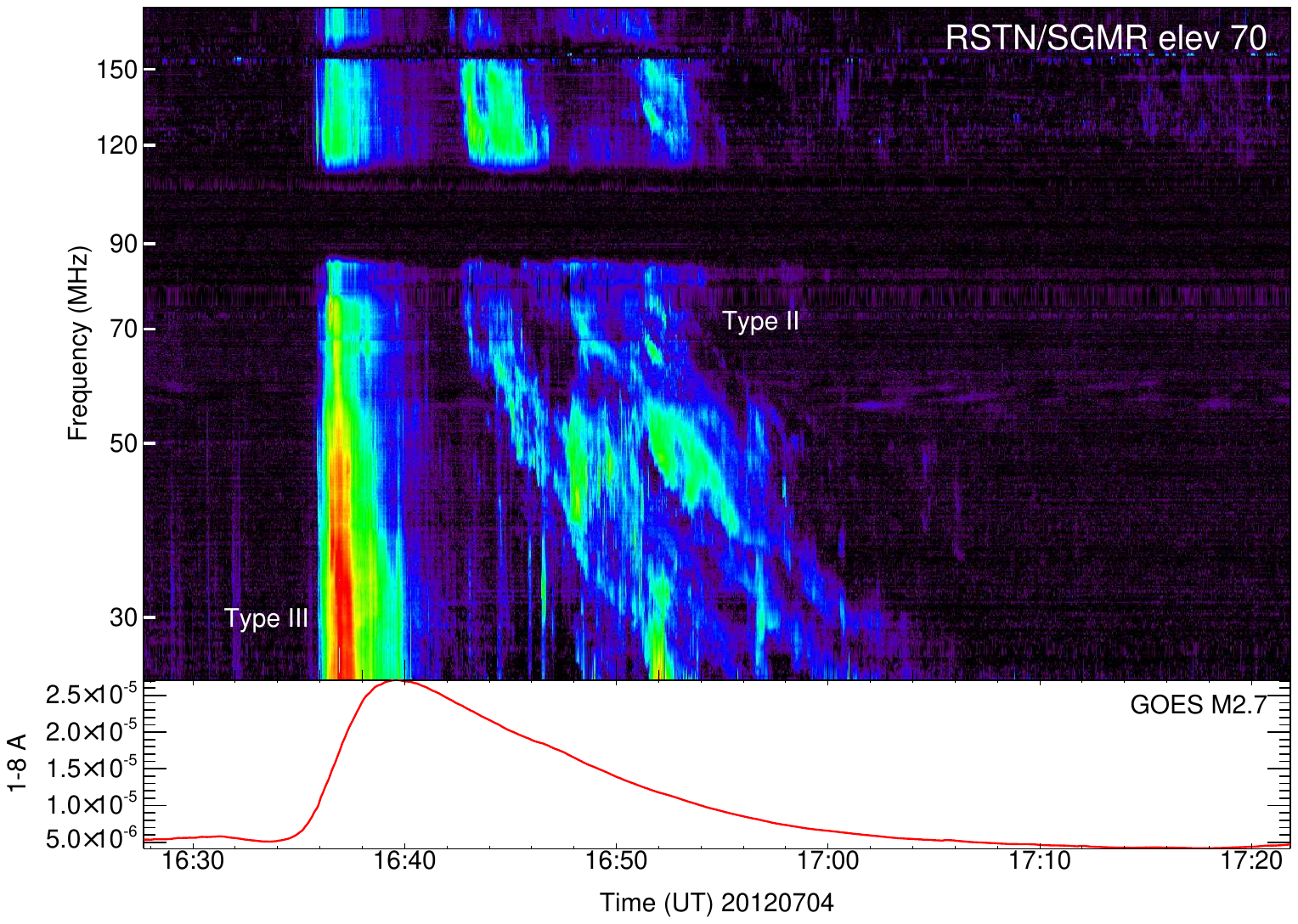}
\caption{This burst from an M2.7 flare on 2012 July 4 exhibits a bright Type III burst in the impulsive phase together with a Type II burst with very wide bands during the decay phase. \label{fig23}}
\end{figure}

\begin{itemize}

\item {\bf Type II bursts} (metric, 25-180 MHz): these show up in a frequency-time plot (``dynamic spectrum'')
as relatively narrowband features drifting towards lower frequencies at rates of order 1 MHz/s or less (with the frequency drift rate slowing as frequency decreases). They often have both
fundamental and second-harmonic branches (i.e., features with a frequency ratio of close to 2:1), and each of these branches can exhibit further splitting in frequency.
For standard coronal density-versus-height models, these drift rates correspond to speeds in the
500-2000 km/s range, i.e., around the expected Alfv{\'e}n speed in the corona, and therefore it is 
assumed that they are produced by shocks. They usually occur after the impulsive phase
of the flare, suggesting that the shock forms after a disturbance propagates to a region
of lower Alfv{\'e}n speed \citep{GLK01}. While CMEs would seem to provide a natural driver for the shock producing metric Type IIs, they often occur in weak flares without a detectable CME \citep[see references cited above, and e.g.][]{Whi07,MMZ12,MPK23,KuG25}. Further, when there is a CME, imaging often shows that the metric Type II radio emission is not co-located with the CME \citep[e.g.,][]{GDH84,Pic99}.

\item {\bf Type III bursts} (metric, impulsive phase): bursts with much faster frequency-time drift rates,
corresponding to the motion of electrons with energies of keV and larger \citep[e.g.][]{LPG81,Lin06}. They often
occur early in the rise of the
impulsive phase. They result from the emission of Langmuir waves by electron beams on 
open field lines. They also may have fundamental and harmonic traces, and in that case the fundamental trace often consists of discrete narrowband features called striae, whereas the second harmonic trace is generally smooth in both frequency and time axes.

\begin{figure}[t]
\centering
\includegraphics[scale=0.62,keepaspectratio=true,angle=0,clip=false]{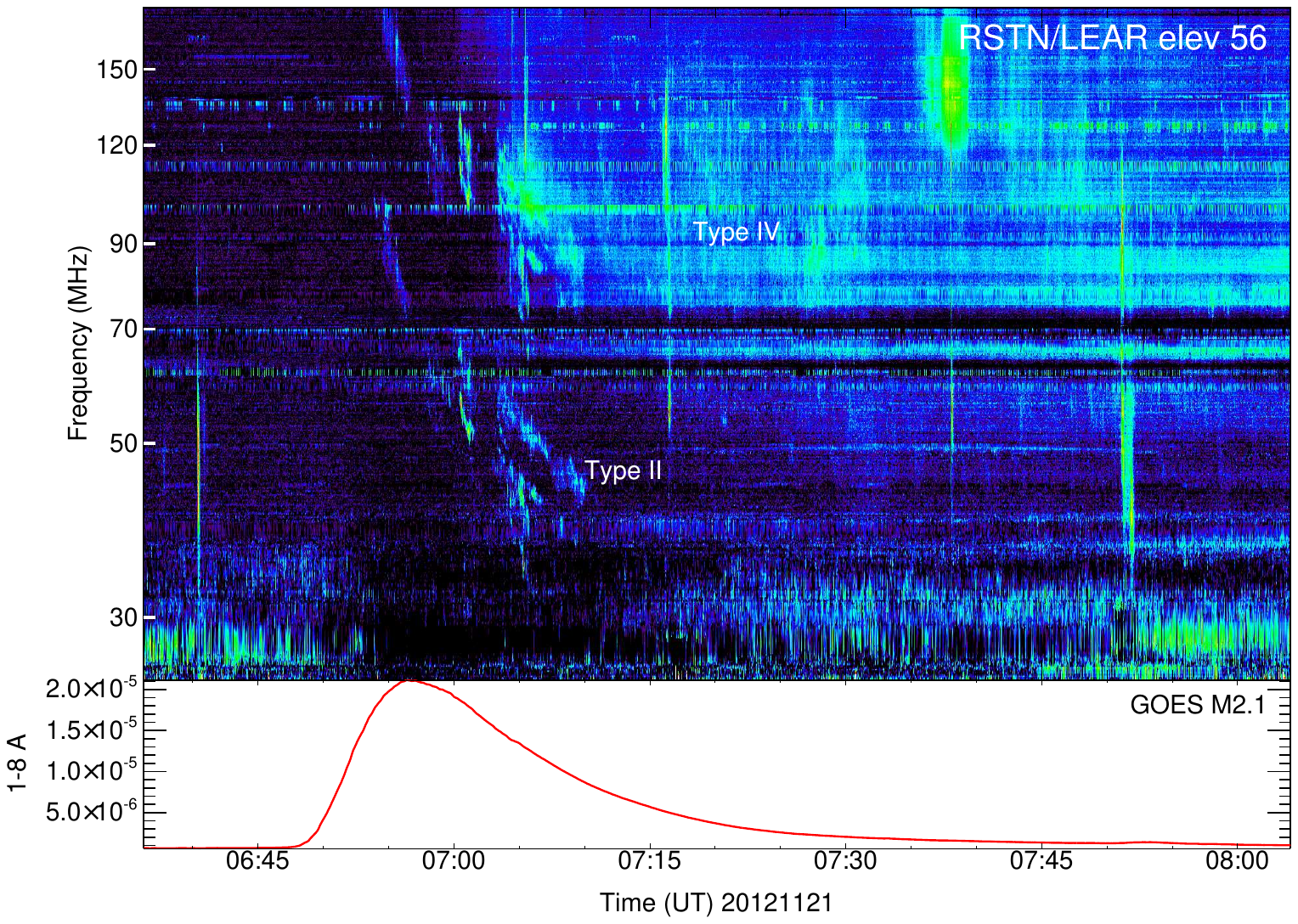}
\caption{In this M2.1 flare from 2012 November 21, narrow fundamental and harmonic traces of a Type II burst can be seen starting close to the peak of the soft X-ray emission, and it is then followed by a Type IV burst, again most prominent above 100 MHz. There is no impulsive phase Type III burst in this event, and a ``short-wave fadeout'' due to excess ionization and absorption in the ionospheric D layer due to the flare soft X-rays is apparent at the lowest frequencies. \label{fig24}}
\end{figure}

\item {\bf Type IV bursts}: these are relatively broad-band, typically low frequency bursts
that occur in the decay phase of the flare and can last for an hour or longer. There are
believed to be two forms of Type IV: ``stationary'' (on the Sun) bursts that are
generally thought to be associated with post-flare loop systems; and ``moving'' Type IVs
thought possibly to be moving outwards in conjunction with a CME \citep[e.g.][]{BPK01}. This distinction is ignored
here since it requires high-quality low-frequency imaging data that is not available for the
vast majority of the cataloged flares. Type IV bursts are thought usually to be preceded by Type II
bursts, as in the example in Fig.~4 \citep[e.g.,][]{CaR88b}.

\item {\bf ``Late'' Type III bursts} (Type III-L or ``T3L''): (metric) Type IIIs that occur typically  as a group of
bursts after the impulsive phase (here taken to be after the peak in soft X-rays). Again, they
require open field lines: they  have been correlated with the occurrence of SEPs, suggesting
an association with CMEs, but this is not a resolved question \citep{CEP02,ClL09,GoM10,DWR15,NuP20,PMK24}. T3Ls were initially thought to be shock-accelerated \citep{CSF81,DLB00}, but later it became clear that they did not originate at the metric Type II shock \citep{CEP02}. The connection with SEPs is usually in the context of long-duration low frequency emission, rather than the metric IIIs that we address here: e.g., \citet{ClL09} investigate Type III emission at 1 MHz and find that the Type III bursts associated with gradual SEP events generally have durations longer than for impulsive SEP events. We do not impose a duration requirement on T3Ls here.  T3Ls do seem to be a common
occurrence that is quite distinct from the impulsive-phase Type IIIs, which are associated
with the onset of flare energy release and generally appear to be single bright bursts. Since the decay of a flare takes much longer than
the impulsive phase, there can be confusion in identification between a T3L associated with a flare and an unrelated III from a different region, so there could certainly be some inflation in the T3L numbers, particularly for eruptive flares which tend to have decay times longer than those of confined flares.

\item {\bf Interplanetary Type II bursts} (IP Type IIs, also known as dekametric/hectometric,
or DH, Type IIs): these occur at frequencies below 10
MHz and need to be observed from space due to the ionospheric cutoff. \citet{CaE05} made a distinction between bursts that appear to be straightforward extensions of metric ($>$ 25 MHz) coronal Type II bursts into the range below 14 MHz, and a separate class of bursts (that they specifically identified as ``IP Type IIs'') that start at lower frequencies and are not connected to metric Type IIs. The latter bursts tend to have smoother frequency and time profiles than the classic plasma-emission metric Type IIs, which led to the suggestion that they may be due to synchrotron emission \citep[e.g.][]{Bas07,PAV13,Poh25}. IP Type IIs tend to have a stronger association with CMEs than do metric Type IIs.

\end{itemize}

\noindent Figures 1-4 show examples of events in which different combinations of metric burst types occurred. In two of the cases Type IV bursts are seen even though the corresponding X-ray flares would not be regarded as long duration. Note that these examples were chosen for clarity, and many of the dynamic spectra used for this study suffer from higher noise levels and worse contamination from radio-frequency interference (RFI) than the plotted examples might suggest.

There is necessarily significant subjectivity in classifying solar radio burst types from dynamic spectra. Individual Type III bursts are usually straightforward thanks to their high frequency drift rates, but an intense cluster of overlapping Type IIIs could be confused with long-lived continuum emission. Type II bursts are the most difficult to identify unambiguously: when present as narrowband harmonic pairs drifting over a wide frequency range, they are straightforward, but when the traces have an instantaneously broad frequency coverage with a lot of structure, when either the fundamental or harmonic trace is missing, or if the burst duration is very short (a few minutes or less), they may not be clear. The presence of harmonic structure in slowly drifting bursts is usually regarded as the most reliable indicator of Type II emission.

It has long been argued \citep[e.g.,][]{CMR86,CaR88a} that short-duration (``impulsive'',
sometimes taken to mean ``confined'') flares are more likely to be associated with 
impulsive-phase Type IIIs, while longer-duration flares (which often seem to be eruptive)
are more likely to have Type II and IV bursts. \citet{Kaz23} indeed found that confined flares tend to have a faster rise and shorter duration than eruptive flares.

\section{Radio bursts in eruptive and confined flares}

The eruptive-confined flare catalog contains 480 flares, of which 152 are eruptive and 
328 are confined. To determine radio burst associations, the following was carried out:

\begin{itemize}

\item For 455 flares, 25-180 MHz data from the USAF Radio Solar Telescope 
Network (RSTN) is available in the NESDIS\footnote{National Environmental Satellite, Data, and Information Service, operated by the National Oceanic and Atmospheric Administration.} archive of RSTN data. The timing in the flare
catalog was used to generate suitable plotting commands for dynamic spectra
covering the event, typically out to an hour from onset in order to be able to see Type
IV bursts, but sometimes longer in the case of very long duration flares. The GOES 1-8 \AA\ light curve is plotted with the radio data in order to
establish timing relative to the SXR onset and flare peak. Figs. 1-4 are examples of such plots. Bursts were identified and
classified visually. In many cases flares were seen by 2 of the 4 RSTN sites, in which case both data sets were inspected (totaling over 700 dynamic spectra).

\item Of the remaining 25 flares, 8 were observed by the Culgoora radio spectrograph in
Australia. Dynamic spectra were generated for these events and analyzed.

\item For the remaining 17 events, the burst reports in the NOAA/SWPC daily event files
were used. These reports are produced by USAF observers at the RSTN sites, based on real-time inspection of the dynamic spectra.

\end{itemize}

\begin{deluxetable}{cccccc}
\tablecaption{Full Eruptive/Confined Catalog}
\tablecolumns{6}
\tablewidth{0pc}
\tablehead{\colhead{} & \colhead{Total flares} & \colhead{Type II} & \colhead{Impulsive Type III} & \colhead{Type IV} & \colhead{Late Type III} }
\startdata
Eruptive & 151 & 52 (34\%) & 89 (59\%) & 70 (46\%) & 66 (44\%) \\
Confined & 323 & 6 (2\%) & 56 (17\%) & 13 (4\%) & 50 (15\%) \\
\enddata
\end{deluxetable}

Association of radio bursts in the 25-180 MHz range was therefore possible for essentially all 480 flares
in the catalog. The visual burst classification was carried out without consulting the catalog,
i.e., without any knowledge of whether a given flare had been labelled eruptive or
confined. It was also done without referring to the original classifications by RSTN
observers. Subsequently, a check was made for flares for which RSTN observers 
reported a Type II burst that we did not identify in the dynamic spectrum:
there were 6 such events, and those dynamic
spectra were re-inspected for completeness, but in each case we could not see clear evidence for a Type II burst.

The results for the metric radio bursts are shown in Table 1\footnote{In the following we exclude from the analysis 6 events (1 eruptive, 5 confined) that occurred during a LASCO data gap in 2012 May for which the presence or absence of a CME cannot be confirmed.}. Several results are striking: 

\begin{itemize}
\item All radio burst types are much more likely to occur in eruptive flares than in
confined flares, and Type II and IV bursts in particular have a very low degree of association with confined flares.
\item The low percentage of confined flares with impulsive-phase Type IIIs seems surprising given the previous analyses. 
\item It is interesting that of the 84 events (35 eruptive, 49 confined) which have an impulsive-phase Type III burst but no Type II or Type IV burst, none are X flares (on either the legacy or the modern scale). 
\item The fact that 10\% of the 58 Type II bursts found across 474 flares had no detectable CME is consistent with the study of 1203 metric Type II bursts by \citet{KMK23}, which found that 15\% of Type II bursts had no associated CME. 
\item Similarly, \citet{KMK21} found that 19\% of the metric Type IVs in their sample had no associated CME: here we find that 16\% of the Type IVs found in the full sample occurred in confined events with no CME. 
\item It is notable, and again surprising, that more Type IV bursts than Type II bursts were found 
in the eruptive flare group: in fact, only 38 of the 70 eruptive flares with Type IVs 
also have Type II bursts. In their survey of 657 Type II and 227 Type IV bursts, \citet{CaR88b} found that only 12\% of the Type IV bursts were not accompanied by Type IIs, and only 30\% of Type II bursts were accompanied by Type IV bursts.
\end{itemize}

\noindent One caution for this last result is that Type IV bursts can easily be confused with ongoing
noise storms, which are broadband features (actually more common at 300 MHz than at 100
MHz) produced by vigorous active regions that are not flare-associated.
When looking at a limited time period around a flare, it can be unclear
if long-lasting continuum emission is Type IV or noise storm. The number of Type IVs
without Type IIs was somewhat surprising, so those cases were also re-checked, and it was determined that 10 of the events identified as Type IVs in the initial inspection of the dynamic spectra were more likely ongoing noise storm emission. The removal of these 10 events still left the 70 Type IV bursts shown in Table 1.

\begin{figure}
\includegraphics[scale=0.65,keepaspectratio=true,clip=false]{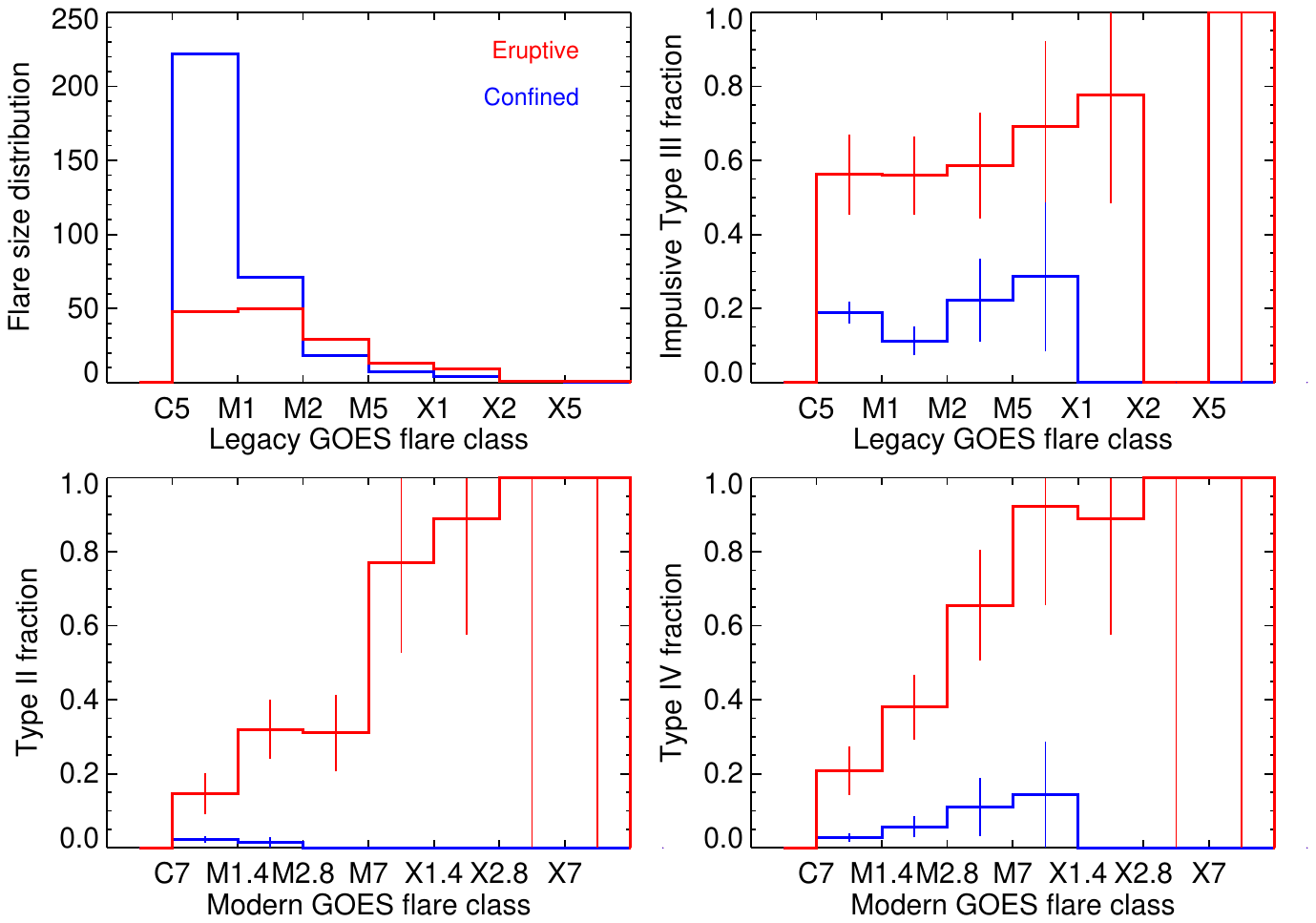}
\caption{Properties of the radio burst associations with eruptive and confined flares as a function of flare size (peak soft X-ray flux). The histogram bins increase by factors of 2 in flare size from left to right. The top left panel plots the distribution of flare sizes (numbers of flares) between the two flare classes (eruptive = red, confined = blue), showing that many more of the confined flares are smaller whereas the larger flares tend to be eruptive. The remaining panels show the fraction of flares in each bin that exhibit the appropriate radio burst type. The top right panel shows how the incidence of Type II bursts changes with flare size for the two flare classes; the lower panels repeat this plot for impulsive Type IIIs (left) and Type IVs (bottom). Error bars are plotted at $\pm\,1\,\sigma$. The upper panels label the flare size bins on the legacy scale, while the lower panels use the modern scale, for the same flare bins. \label{figfrac}}
\end{figure}

It is noteworthy however that on average the eruptive flares in the catalog are much
larger than the confined flares in the sense of peak SXR flux. Figure \ref{figfrac} plots the distribution of flare sizes for the eruptive and confined classes: the median size of a confined flare is M1.1 on the modern scale, while the median eruptive flare is M2.1. This will certainly affect
associations with radio bursts, since the bigger a flare is, the more likely it is to
have a wide range of detectable phenomena \citep[e.g.][]{Kah82a}. The histograms of SXR flare peak sizes for the
two classes of flare look relatively similar above GOES class M1. To try to minimize 
the size bias, we follow \citet{Kaz23} and present in Table 2 the results for the 204 flares in the catalog at M1.0 (on the legacy GOES XRS scale) and larger, of which 103 are eruptive and 101 are confined. The
associations with confined flares don't actually change significantly in any category,
whereas the percentages for eruptive flares all increase significantly when smaller flares are excluded. The remaining 3 panels in Fig. \ref{figfrac} show how the incidence of radio burst types changes with flare size, and emphasizes the contrast between eruptive and confined flares. Interestingly, while the occurrence of impulsive-phase Type III bursts does not change greatly with flare size, for either eruptive or confined flares, it does increase steeply with flare size for Type II and IV bursts in association with eruptive flares (bearing in mind the fact that there are relatively few X flares in the sample).

\begin{deluxetable}{cccccc}
\tablecaption{Eruptive/Confined $>$M1}
\tablecolumns{6}
\tablewidth{0pc}
\tablehead{\colhead{} & \colhead{Total flares} & \colhead{Type II} & \colhead{Impulsive Type III} & \colhead{Type IV} & \colhead{Late Type III} }
\startdata
Eruptive & 103 & 45 (44\%) & 62 (60\%) & 60 (58\%) & 50 (49\%) \\
Confined & 101 & 1 (1\%) & 14 (14\%) & 7 (7\%) & 15 (15\%) \\
\enddata
\end{deluxetable}

For the IP Type II bursts, the list maintained by the CDAW group at
NASA/GSFC, derived from the WAVES instruments on the WIND and STEREO spacecraft, was
used \citep[\url{http://cdaw.gsfc.nasa.gov/CME_list/radio/waves_type2.html}, see][]{GMY19}. The list was
inspected for events that matched the flares in the catalog, and 35 matches were found. IP Type IIs
can last for many hours, and thus specific associations can be difficult to determine
(usually best done from the timing of bright Type IIIs that usually precede IP Type IIs
at DH frequencies).
Not surprisingly in view of the results above, all 35 flares associated with IP Type IIs are
eruptive. Perhaps more surprising is the fact that only 24 of the 35 have metric Type IIs
as well, while 29 of the 35 also showed metric Type IV emission. Most of the flares
producing IP Type IIs are large: there are only 2 C-class flares (on the legacy GOES scale) 
among the 35.  As mentioned above, there 
seem to be two classes of IP Type II: those that are clearly just extensions of
the metric Type IIs from higher frequencies, which usually don't propagate much below 
10 MHz; and bursts that start at a 
frequency below 10 MHz but can last for a long time (hours), and are
likely driven directly by the IP CME. There is no effort to make that distinction in the CDAW catalog, 
and inspection of the burst plots shows that both are present in our sample: we judge that 14 of the 35 matching ``IP Type IIs'' in the CDAW catalog do not radiate much below 10 MHz, and all but 3 of these 14 occur in conjunction with a metric Type II.

\section{Discussion}

The issue of why some flares produce CMEs and others do not is clearly important for understanding the solar flare mechanism(s): is there just one mechanism, in which case apparently an eruption is not required for a flare to occur, or are there actually two different mechanisms operating? One might wish to get around the first option by assuming that all confined flares are actually events with failed eruptions. Such an explanation would seem to be consistent with the studies of the magnetic properties of active regions producing eruptive and confined flares. \citet{LCH21} found that the fraction of flares from a given active region that are eruptive decreases greatly as the total magnetic flux of the region increases \citep[see also][]{BTV18,LSH22,GTV24,LRW24}. This result is attributed to the confinement of coronal material by overlying magnetic fields. More care is needed to identify failed eruptions against the background of the solar disk in EUV images than to see CMEs in coronagraph images, but there are enough well-observed flares with no evidence of a role for filament motion or any other form of eruptive behavior that it seems unlikely that all confined flares are failed eruptions. \citet{DGY25} found that only 1\% of the 1225 prominence eruptions in their study failed, while \citet{MKS20} carried out an automated survey of AIA EUV images from 2010 to 2019 and found twice as many successful eruptions as failed eruptions, which is not consistent with the fact that there are more confined flares than eruptive.

In the survey presented here we have found stark differences in the incidence of radio bursts in eruptive and confined flares: eruptive flares are much more likely to exhibit all types of classic metric radio bursts.  The results of our survey of bursts below 200 MHz complement and are consistent with the study of microwave emission above 400 MHz carried out by \citet{CKH25}. They investigated 29 confined flares from a single active region (AR 12192 in 2014 October), and showed that the confined flares had very little radio emission below 1 GHz, and were unlikely to have emission at 1 MHz indicative of Type III electron beams escaping the flare region \citep[consistent with the results of][mentioned above]{Mit21}. In contrast, $>$M5 eruptive flares from \citet{Kaz23} were far more likely than $>$M5 confined flares to exhibit radio bursts at 410 MHz, and generally produced interplanetary Type III bursts.  \citet{CKH25} interpret these results to mean that confined flares do not usually involve open field lines, ruling out interchange reconnection \citep{CGK02}, i.e., reconnection
between open and closed field lines, as a mechanism for such events, but consistent with reconnection between closed loops. Interchange reconnection propels mass ballistically outward along open field
lines, a type of mass ejection termed a jet. SEP events associated with jets are
characterized by their high electron-to-proton, $^3$He/$^4$He, and Fe/O ratios \citep{Rea99a}.
Jets are qualitatively different from the closed field eruptions of CMEs (although they
may be classified as such in CME catalogs) and cannot drive a coronal shock because
the plasma is not moving perpendicular to the field \citep{VrC08}. Because jets
are formed by reconnection involving an open field line, they can naturally produce type III bursts due to plasma emission excited by outward flowing electrons.

The frequent occurrence of bright Type III bursts early in the impulsive rise phase of flares is well known, but has not received a lot of theoretical attention. There are two aspects to this phenomenon: it seems to imply that there is a significant role for open field lines in flare onset, and that the generation of an outwards-propagating electron beam on the open field lines is often a key part of the flare trigger. It is important to note that the Type III electron beam is distinct from the nonthermal accelerated electrons responsible for flare hard X-ray emission from chromospheric footpoints, and requires a separate acceleration episode: the impulsive Type III beam is generally a single feature of much shorter duration than flare hard X-rays. The standard cartoon for flares in a single bipole with a strongly sheared neutral line \citep[e.g.][]{MSH01,MTP24} has no open field lines until the CME blows out the corona above the active region. Since neither Type II nor Type IV bursts have any intrinsic association with open field lines, the higher association (45-60\%) of these bursts, as well as metric and DH Type III bursts, with eruptive events requires further explanation. If stationary Type IV bursts are indeed linked to post-flare loops, then their association is explicitly with closed field lines. However, \citet{SaK20} find that stationary Type IV bursts are not on flare loops, but rather are located in one leg of the flux rope forming the core of the CME, implying that they are on very long field lines as a CME moves outwards over the typically hour-long duration of a Type IV event.

One possibility is that quasi-perpendicular shocks driven by
the lateral expansion of a CME accelerate Type III electrons on open field lines adjacent
to the CME. Quasi-perpendicular shocks have been linked to electron acceleration in IP
shocks at Earth \citep{TsL85}. For a sample of 169 large (width $>$ 140\arcdeg) CMEs, Cane
et al. (2002) noted that 66 of these flares (40\%) lacked any Type III emission. Of these
66 eruptive flares, 56 had CME speeds $<$ 600 km s$^{-1}$, hinting that Type III emission at the
onset of classic fully-developed radio bursts \citep[see Figure 1 in ][]{Dul85} may
result from such quasi-perpendicular shock acceleration of electrons driven by the
lateral expansion of a CME, reversing the normal classical order of the radio emissions
(with the delayed appearance of type II emission attributed to a CME-driven bow
shock). The 45\% of eruptive flares in our sample with associated early Type III
emission had a median CME speed of 660 km s$^{-1}$ (using speeds given in the CDAW CME catalog), compared to 373 km s$^{-1}$ for those lacking such
emission (albeit with larger median flare sizes: M3.3 with a Type III, M2.7 without). The eruptive flares in Table 2 could also
have associated type III emission if they begin as a standard jet and evolve to a
blow-out type jet with an associated CME \citep{MCS10,CZL15}.
For eruptive flares, late-phase type III emission could result from reconnection at the
X-point or in the neutral current sheet created by the CME. Late Type III associations
might also occur by chance due to bursts unrelated to the flare under consideration.
For confined flares, early metric Type III emission could be due to a surge (a type
of jet for which the plasma does not escape from the Sun \citep[see ][p. 221]{Sve76},
caused by reconnection between a small loop and a large loop that does not evolve into
a CME \citep[see examples in the movies in Figures 2(a,b) from ][]{CZL15}. Such surges
show evidence of intense emission at 245 MHz that is not present at 1 MHz, indicating
the absence of electrons that escape from the Sun \citep{CKH25}.

We should note that the results here are specific to the sample in the \citet{Kaz23} catalog and analysis of a wider sample of eruptive and confined events would be valuable as a check. In principle the limit to events within 45\arcdeg\ of central meridian imposed in the catalog could produce a bias in our association results, but metric solar radio bursts are not believed to be strongly beamed, and are readily detected from flares at and often beyond the solar limb, so we think it unlikely that this restriction plays a role in the outcomes.

\section{Summary}

We find that: (a) Large (>M1.4) confined flares characteristically are deficient in
metric Type II, III, and IV emission, consistent with an origin in the reconnection of closed loops, with associations for all three types in the range of ~1-15\%; and (b) eruptive
(CME-associated) flares have high (45-60\%) degrees of association with metric Type II, III, and IV
bursts. Because reconnection of closed loops does not involve pre-existing open field lines, we
suggest that Type III emission in eruptive flares may arise from: (i) quasi-perpendicular
shock acceleration of electrons on open field lines on the flanks of a CME; and (ii) instances
when a standard jet-type flare evolves to a blow-out jet for impulsive-phase type IIIs. The frequent occurrence of Type III bursts early in the rise of the flare impulsive phase suggests that the mechanism that produces the Type III plays a significant role in eruptive flare onset.
We interpret the above results in terms of the three basic types of magnetic
reconnection topologies for flares: (a) loop-loop interactions (closed-closed) for confined
flares in which neither plasma nor particles escape from the Sun; (b) loop and open
field interactions (closed-open) giving rise to jets and Type III emission; and X-point
reconnection between oppositely oppositely-directed field lines in a loop that creates
two closed field structures, one above (a CME that drives Type II shock waves) and
one below (flare loops) the X-point.

\acknowledgments
SW thanks AFOSR for support for basic research through LRIR 26RVCOR010. The views expressed herein are those of the authors and do not reflect official guidance of the US Government or Dept. of Defense. The appearance of external hyperlinks does not constitute endorsement by the US Dept. of Defense of the contents of such links. 

\facilities{This work uses data from the US Air Force Radio Solar Telescope Network.}
\vspace{\baselineskip}

{\large\it Data availability:} RSTN solar radio spectrograph (SRS) data are available from \url{https://www.ngdc.noaa.gov/stp/space-weather/solar-data/solar-features/solar-radio/rstn-spectral/}. Data from the Culgoora radio spectrograph operated by the Australian Bureau of Meteorology are available at \url{https://downloads.sws.bom.gov.au/wdc/wdc_spec/data/culgoora/raw/}. The 757 dynamic spectra used to classify the radio burst types and updated spreadsheets of eruptive and confined flare properties are available in the Zenodo archive (doi:10.5281/zenodo.20318255).

\clearpage

\bibliographystyle{aasjournal}
%\bibliography{solar}

\begin{thebibliography}{}
\expandafter\ifx\csname natexlab\endcsname\relax\def\natexlab#1{#1}\fi
\providecommand{\url}[1]{\href{#1}{#1}}
\providecommand{\dodoi}[1]{doi:~\href{http://doi.org/#1}{\nolinkurl{#1}}}
\providecommand{\doeprint}[1]{\href{http://ascl.net/#1}{\nolinkurl{http://ascl.net/#1}}}
\providecommand{\doarXiv}[1]{\href{https://arxiv.org/abs/#1}{\nolinkurl{https://arxiv.org/abs/#1}}}

\bibitem[{{Bastian}(2007)}]{Bas07}
{Bastian}, T.~S. 2007, Astrophys. J., 665, 805, \dodoi{10.1086/519246}

\bibitem[{Bastian {et~al.}(2001)Bastian, Pick, Kerdraon, Maia, \& Vourlidas}]{BPK01}
Bastian, T.~S., Pick, M., Kerdraon, A., Maia, D., \& Vourlidas, A. 2001, Astrophys. J. Letters, 558, 65

\bibitem[{{Baumgartner} {et~al.}(2018){Baumgartner}, {Thalmann}, \& {Veronig}}]{BTV18}
{Baumgartner}, C., {Thalmann}, J.~K., \& {Veronig}, A.~M. 2018, Astrophys. J., 853, 105, \dodoi{10.3847/1538-4357/aaa243}

\bibitem[{{Cane} \& {Erickson}(2005)}]{CaE05}
{Cane}, H.~V., \& {Erickson}, W.~C. 2005, Astrophys. J., 623, 1180, \dodoi{10.1086/428820}

\bibitem[{Cane {et~al.}(2002)Cane, Erickson, \& Prestage}]{CEP02}
Cane, H.~V., Erickson, W.~C., \& Prestage, N.~P. 2002, J. Geophys. Res., 107, 1315, \dodoi{10.1029/2001JA000320}

\bibitem[{{Cane} {et~al.}(1986){Cane}, {McGuire}, \& {von Rosenvinge}}]{CMR86}
{Cane}, H.~V., {McGuire}, R.~E., \& {von Rosenvinge}, T.~T. 1986, Astrophys. J., 301, 448, \dodoi{10.1086/163913}

\bibitem[{Cane \& Reames(1988{\natexlab{a}})}]{CaR88b}
Cane, H.~V., \& Reames, D.~V. 1988{\natexlab{a}}, Astrophys. J., 325, 901, \dodoi{10.1086/166061}

\bibitem[{Cane \& Reames(1988{\natexlab{b}})}]{CaR88a}
---. 1988{\natexlab{b}}, Astrophys. J., 325, 895, \dodoi{10.1086/166060}

\bibitem[{Cane {et~al.}(1981)Cane, Stone, Fainberg, Stewart, \& Steinberg}]{CSF81}
Cane, H.~V., Stone, R.~G., Fainberg, J., Stewart, R.~T., \& Steinberg, J.~L. 1981, Geophys. Res. Lett., 8, 1285, \dodoi{10.1029/GL008i012p01285}

\bibitem[{Carmichael(1964)}]{Car64}
Carmichael, H. 1964, in AAS-NASA Symposium on Solar Flares, ed. W.~N. Hess (NASA SP-50), 451

\bibitem[{{Chen} {et~al.}(2015){Chen}, {Zhang}, {Ma}, {Yang}, {Li}, {Huang}, \& {Xiao}}]{CZL15}
{Chen}, H., {Zhang}, J., {Ma}, S., {et~al.} 2015, Astrophys. J. Letters, 808, L24, \dodoi{10.1088/2041-8205/808/1/L24}

\bibitem[{{Cliver} {et~al.}(2025){Cliver}, {Kazachenko}, {Hudson}, {Alberti}, {Laurenza}, {White}, \& {Gallagher}}]{CKH25}
{Cliver}, E.~W., {Kazachenko}, M., {Hudson}, H.~S., {et~al.} 2025, Astrophys. J., 994, 103, \dodoi{10.3847/1538-4357/adfbe5}

\bibitem[{{Cliver} \& {Ling}(2009)}]{ClL09}
{Cliver}, E.~W., \& {Ling}, A.~G. 2009, Astrophys. J., 690, 598, \dodoi{10.1088/0004-637X/690/1/598}

\bibitem[{{Crooker} {et~al.}(2002){Crooker}, {Gosling}, \& {Kahler}}]{CGK02}
{Crooker}, N.~U., {Gosling}, J.~T., \& {Kahler}, S.~W. 2002, Journal of Geophysical Research (Space Physics), 107, 1028, \dodoi{10.1029/2001JA000236}

\bibitem[{{DeRosa} \& {Barnes}(2018)}]{DeB18}
{DeRosa}, M.~L., \& {Barnes}, G. 2018, Astrophys. J., 861, 131, \dodoi{10.3847/1538-4357/aac77a}

\bibitem[{{Devi} {et~al.}(2025){Devi}, {Gopalswamy}, {Yashiro}, {Akiyama}, {Chandra}, \& {Koleva}}]{DGY25}
{Devi}, P., {Gopalswamy}, N., {Yashiro}, S., {et~al.} 2025, Journal of Astrophysics and Astronomy, 46, 64, \dodoi{10.1007/s12036-025-10088-2}

\bibitem[{{Duffin} {et~al.}(2015){Duffin}, {White}, {Ray}, \& {Kaiser}}]{DWR15}
{Duffin}, R.~T., {White}, S.~M., {Ray}, P.~S., \& {Kaiser}, M.~L. 2015, Journal of Physics Conference Series, 642, 012006, \dodoi{10.1088/1742-6596/642/1/012006}

\bibitem[{{Dulk}(1985)}]{Dul85}
{Dulk}, G.~A. 1985, araa, 23, 169, \dodoi{10.1146/annurev.aa.23.090185.001125}

\bibitem[{{Dulk} {et~al.}(2000){Dulk}, {Leblanc}, {Bastian}, \& {Bougeret}}]{DLB00}
{Dulk}, G.~A., {Leblanc}, Y., {Bastian}, T.~S., \& {Bougeret}, J.-L. 2000, J. Geophys. Res., 105, 27343, \dodoi{10.1029/2000JA000076}

\bibitem[{{Gary} {et~al.}(1984){Gary}, {Dulk}, {House}, {Illing}, {Sawyer}, {Wagner}, {McLean}, \& {Hildner}}]{GDH84}
{Gary}, D.~E., {Dulk}, G.~A., {House}, L., {et~al.} 1984, Astron. Astrophys., 134, 222

\bibitem[{{Gopalswamy} {et~al.}(2001){Gopalswamy}, {Lara}, {Kaiser}, \& {Bougeret}}]{GLK01}
{Gopalswamy}, N., {Lara}, A., {Kaiser}, M.~L., \& {Bougeret}, J.-L. 2001, J. Geophys. Res., 106, 25261, \dodoi{10.1029/2000JA004025}

\bibitem[{{Gopalswamy} \& {M{\"a}kel{\"a}}(2010)}]{GoM10}
{Gopalswamy}, N., \& {M{\"a}kel{\"a}}, P. 2010, Astrophys. J. Letters, 721, L62

\bibitem[{{Gopalswamy} {et~al.}(2019){Gopalswamy}, {M{\"a}kel{\"a}}, \& {Yashiro}}]{GMY19}
{Gopalswamy}, N., {M{\"a}kel{\"a}}, P., \& {Yashiro}, S. 2019, Sun and Geosphere, 14, 111, \dodoi{10.31401/SunGeo.2019.02.03}

\bibitem[{{Gupta} {et~al.}(2024){Gupta}, {Thalmann}, \& {Veronig}}]{GTV24}
{Gupta}, M., {Thalmann}, J.~K., \& {Veronig}, A.~M. 2024, Astron. Astrophys., 686, A115, \dodoi{10.1051/0004-6361/202346212}

\bibitem[{Hirayama(1974)}]{Hir74}
Hirayama, T. 1974, Solar Phys., 34, 323

\bibitem[{Kahler(1982)}]{Kah82a}
Kahler, S.~W. 1982, J. Geophys. Res., 87, 3439

\bibitem[{{Kazachenko}(2023)}]{Kaz23}
{Kazachenko}, M.~D. 2023, Astrophys. J., 958, 104, \dodoi{10.3847/1538-4357/ad004e}

\bibitem[{Kopp \& Pneuman(1976)}]{KoP76}
Kopp, R.~A., \& Pneuman, G.~W. 1976, Solar Phys., 50, 85

\bibitem[{{Kumari} \& {Gopalswamy}(2025)}]{KuG25}
{Kumari}, A., \& {Gopalswamy}, N. 2025, Journal of Astrophysics and Astronomy, 46, 90, \dodoi{10.1007/s12036-025-10115-2}

\bibitem[{{Kumari} {et~al.}(2021){Kumari}, {Morosan}, \& {Kilpua}}]{KMK21}
{Kumari}, A., {Morosan}, D.~E., \& {Kilpua}, E.~K.~J. 2021, Astrophys. J., 906, 79, \dodoi{10.3847/1538-4357/abc878}

\bibitem[{{Kumari} {et~al.}(2023){Kumari}, {Morosan}, {Kilpua}, \& {Daei}}]{KMK23}
{Kumari}, A., {Morosan}, D.~E., {Kilpua}, E.~K.~J., \& {Daei}, F. 2023, Astron. Astrophys., 675, A102, \dodoi{10.1051/0004-6361/202244015}

\bibitem[{Kundu(1965)}]{Kun65}
Kundu, M.~R. 1965, Solar Radio Astronomy (New York: Interscience Publishers)

\bibitem[{{Lawrance} {et~al.}(2024){Lawrance}, {Devi}, {Chandra}, \& {Miteva}}]{LDC24}
{Lawrance}, B., {Devi}, P., {Chandra}, R., \& {Miteva}, R. 2024, Solar Phys., 299, 75, \dodoi{10.1007/s11207-024-02317-8}

\bibitem[{{Li} {et~al.}(2024){Li}, {Rao}, {Wang}, {Zhao}, {Xiang}, {Deng}, {Li}, \& {Liu}}]{LRW24}
{Li}, F., {Rao}, C., {Wang}, H., {et~al.} 2024, Astrophys. J. Letters, 976, L2, \dodoi{10.3847/2041-8213/ad8c37}

\bibitem[{{Li} {et~al.}(2021){Li}, {Chen}, {Hou}, {Veronig}, {Yang}, \& {Zhang}}]{LCH21}
{Li}, T., {Chen}, A., {Hou}, Y., {et~al.} 2021, Astrophys. J. Letters, 917, L29, \dodoi{10.3847/2041-8213/ac1a15}

\bibitem[{{Li} {et~al.}(2022){Li}, {Sun}, {Hou}, {Chen}, {Yang}, \& {Zhang}}]{LSH22}
{Li}, T., {Sun}, X., {Hou}, Y., {et~al.} 2022, Astrophys. J. Letters, 926, L14, \dodoi{10.3847/2041-8213/ac5251}

\bibitem[{{Lin}(2006)}]{Lin06}
{Lin}, R.~P. 2006, Geophysical Monograph Series, 165, 199, \dodoi{10.1002/9781118666203.ch18}

\bibitem[{{Lin} {et~al.}(1981){Lin}, {Potter}, {Gurnett}, \& {Scarf}}]{LPG81}
{Lin}, R.~P., {Potter}, D.~W., {Gurnett}, D.~A., \& {Scarf}, F.~L. 1981, Astrophys. J., 251, 364, \dodoi{10.1086/159471}

\bibitem[{{Magdaleni{\'c}} {et~al.}(2012){Magdaleni{\'c}}, {Marqu{\'e}}, {Zhukov}, {Vr{\v s}nak}, \& {Veronig}}]{MMZ12}
{Magdaleni{\'c}}, J., {Marqu{\'e}}, C., {Zhukov}, A.~N., {Vr{\v s}nak}, B., \& {Veronig}, A. 2012, Astrophys. J., 746, 152, \dodoi{10.1088/0004-637X/746/2/152}

\bibitem[{McLean \& Labrum(1985)}]{McL85}
McLean, D.~J., \& Labrum, N.~R. 1985, Solar Radiophysics (Cambridge: Cambridge University Press)

\bibitem[{{Miteva}(2021)}]{Mit21}
{Miteva}, R. 2021, Bulgarian Astronomical Journal, 35, 87

\bibitem[{{Miteva} \& {P{\"o}tzi}(2025)}]{MiP25}
{Miteva}, R., \& {P{\"o}tzi}, W. 2025, Bulgarian Astronomical Journal, 43, 34

\bibitem[{{Moore} {et~al.}(2010){Moore}, {Cirtain}, {Sterling}, \& {Falconer}}]{MCS10}
{Moore}, R.~L., {Cirtain}, J.~W., {Sterling}, A.~C., \& {Falconer}, D.~A. 2010, Astrophys. J., 720, 757, \dodoi{10.1088/0004-637X/720/1/757}

\bibitem[{{Moore} {et~al.}(2001){Moore}, {Sterling}, {Hudson}, \& {Lemen}}]{MSH01}
{Moore}, R.~L., {Sterling}, A.~C., {Hudson}, H.~S., \& {Lemen}, J.~R. 2001, Astrophys. J., 552, 833, \dodoi{10.1086/320559}

\bibitem[{{Moore} {et~al.}(2024){Moore}, {Tiwari}, {Panesar}, {Aparna}, \& {Sterling}}]{MTP24}
{Moore}, R.~L., {Tiwari}, S.~K., {Panesar}, N.~K., {Aparna}, V., \& {Sterling}, A.~C. 2024, Astrophys. J., 975, 20, \dodoi{10.3847/1538-4357/ad71d2}

\bibitem[{{Morosan} {et~al.}(2023){Morosan}, {Pomoell}, {Kumari}, {Kilpua}, \& {Vainio}}]{MPK23}
{Morosan}, D.~E., {Pomoell}, J., {Kumari}, A., {Kilpua}, E.~K.~J., \& {Vainio}, R. 2023, Astron. Astrophys., 675, A98, \dodoi{10.1051/0004-6361/202245515}

\bibitem[{{Mrozek} {et~al.}(2020){Mrozek}, {Ko{\l}oma{\'n}ski}, {St{e}{\'s}licki}, \& {Gronkiewicz}}]{MKS20}
{Mrozek}, T., {Ko{\l}oma{\'n}ski}, S., {St{e}{\'s}licki}, M., \& {Gronkiewicz}, D. 2020, Astrophys. J. Supp., 249, 21, \dodoi{10.3847/1538-4365/ab9e00}

\bibitem[{{N{\'u}{\~n}ez} \& {Paul-Pena}(2020)}]{NuP20}
{N{\'u}{\~n}ez}, M., \& {Paul-Pena}, D. 2020, Universe, 6, 161, \dodoi{10.3390/universe6100161}

\bibitem[{{Pick}(1999)}]{Pic99}
{Pick}, M. 1999, in Proceedings of the Nobeyama Symposium, ed. T.~S. {Bastian}, N.~{Gopalswamy}, \& K.~{Shibasaki}, 187--198

\bibitem[{{Pick} \& {Vilmer}(2008)}]{PiV08}
{Pick}, M., \& {Vilmer}, N. 2008, Astron. Astrophys. Rev., 16, 1, \dodoi{10.1007/s00159-008-0013-x}

\bibitem[{{Pohjolainen}(2025)}]{Poh25}
{Pohjolainen}, S. 2025, Solar Phys., 300, 25, \dodoi{10.1007/s11207-025-02449-5}

\bibitem[{{Pohjolainen} {et~al.}(2013){Pohjolainen}, {Allawi}, \& {Valtonen}}]{PAV13}
{Pohjolainen}, S., {Allawi}, H., \& {Valtonen}, E. 2013, Astron. Astrophys., 558, A7, \dodoi{10.1051/0004-6361/201220688}

\bibitem[{{Posner} {et~al.}(2024){Posner}, {Malandraki}, {Karavolos}, {Tziotziou}, {Smanis}, {Heber}, {Dr{\"o}ge}, {K{\"u}hl}, \& {Veldes}}]{PMK24}
{Posner}, A., {Malandraki}, O.~E., {Karavolos}, M., {et~al.} 2024, Space Weather, 22, e2024SW004013, \dodoi{10.1029/2024SW00401310.22541/essoar.171926580.08698491/v1}

\bibitem[{{Reames}(1999)}]{Rea99a}
{Reames}, D.~V. 1999, Space Science Reviews, 90, 413

\bibitem[{{Salas-Matamoros} \& {Klein}(2020)}]{SaK20}
{Salas-Matamoros}, C., \& {Klein}, K.-L. 2020, Astron. Astrophys., 639, A102, \dodoi{10.1051/0004-6361/202037989}

\bibitem[{{Shanmugaraju} {et~al.}(2005){Shanmugaraju}, {Moon}, {Kim}, {Dryer}, \& {Umapathy}}]{SMK05}
{Shanmugaraju}, A., {Moon}, Y.-J., {Kim}, Y.-H., {Dryer}, M., \& {Umapathy}, S. 2005, Solar Phys., 225, 141

\bibitem[{{Shanmugaraju} {et~al.}(2011){Shanmugaraju}, {Moon}, \& {Vr{\v{s}}nak}}]{SMV11}
{Shanmugaraju}, A., {Moon}, Y.-J., \& {Vr{\v{s}}nak}, B. 2011, Solar Phys., 270, 273, \dodoi{10.1007/s11207-011-9752-3}

\bibitem[{{Sturrock}(1966)}]{Stu66}
{Sturrock}, P.~A. 1966, Nature, 211, 695, \dodoi{10.1038/211695a0}

\bibitem[{{Svestka}(1976)}]{Sve76}
{Svestka}, Z. 1976, Solar Flares (Heidelberg: Springer-Verlag Berlin)

\bibitem[{{Tsurutani} \& {Lin}(1985)}]{TsL85}
{Tsurutani}, B.~T., \& {Lin}, R.~P. 1985, J. Geophys. Res., 90, 1, \dodoi{10.1029/JA090iA01p00001}

\bibitem[{{Vr{\v s}nak} \& {Cliver}(2008)}]{VrC08}
{Vr{\v s}nak}, B., \& {Cliver}, E.~W. 2008, Solar Phys., 253, 215, \dodoi{10.1007/s11207-008-9241-5}

\bibitem[{{White}(2007)}]{Whi07}
{White}, S.~M. 2007, Asian Journal of Physics, 16, 189, \dodoi{10.48550/arXiv.2405.00959}

\bibitem[{{Wild} {et~al.}(1963){Wild}, {Smerd}, \& {Weiss}}]{WSW63}
{Wild}, J.~P., {Smerd}, S.~F., \& {Weiss}, A.~A. 1963, Ann. Rev. Astron. Astrophys., 1, 291

\bibitem[{{Yashiro} {et~al.}(2004){Yashiro}, {Gopalswamy}, {Michalek}, {St. Cyr}, {Plunkett}, {Rich}, \& {Howard}}]{YGM04}
{Yashiro}, S., {Gopalswamy}, N., {Michalek}, G., {et~al.} 2004, Journal of Geophysical Research (Space Physics), 109, A07105, \dodoi{10.1029/2003JA010282}

\end{thebibliography}

\end{document}